\documentclass[a4paper,11pt]{article}
\usepackage{PoS} % PoS style
\newcommand{\uvec}{\boldsymbol}
\newcommand{\ud}{\text{d}}
\title{Nucleon 3D intrinsic spin structure from the weak-neutral axial-vector form factors}
\author*[a,b]{Yi Chen}
\affiliation[a]{Department of Physics and Center for High Energy Physics, Tsinghua University, No.~30, Shuangqing Road, Beijing, China}
\affiliation[b]{Interdisciplinary Center for Theoretical Study and Department of Modern Physics, University of Science and Technology of China, No.~96, JinZhai Road, Hefei, China}
\emailAdd{physchen@mail.ustc.edu.cn}
\abstract{Relativistic 3D weak-neutral axial-vector four-current and spin distributions inside a nucleon (or a general spin-$\frac{1}{2}$ hadron) including three weak-neutral axial-vector form factors are investigated for the first time. We clarify that the relativistic 3D axial charge distribution in the Breit frame is completely described by the induced pseudotensor form factor $G_T^Z(Q^2)$ rather than by the axial form factor $G_A^Z(Q^2)$. We demonstrate that the quantity $R_A \equiv \sqrt{ \frac{-6}{G_A^Z(0) }\frac{\text{d} G_A^Z(Q^2) }{\text{d} Q^2} \Big|_{Q^2=0} }$ can not be interpreted as the physically meaningful 3D root-mean-square axial radius of a spin-$\frac{1}{2}$ hadron. The genuine 3D root-mean-square axial radius in fact does not exist for any spin-$\frac{1}{2}$ hadron. We also show that the relativistic 3D weak-neutral spin radius $r_\text{spin} = \sqrt{\langle r_\text{spin}^2 \rangle}$, defined as $\langle r_\text{spin}^2 \rangle \equiv R_A^2 + \frac{ 1 }{ 4M^2 }\left[ 1 + \frac{ 2 G_P^Z(0) }{ G_A^Z(0) } \right]$ based on the relativistic and intrinsic 3D weak-neutral spin distribution in the Breit frame, is a physically meaningful radius that can be unambiguously defined for the nucleon. This provides an additional key motivation for the further determination of the induced pseudoscalar form factor $G_P^Z(Q^2)$, e.g. via lattice QCD or model calculations. Numerically, we find that $R_A \approx 0.6510~\text{fm}$ and $r_\text{spin} \approx 2.1054~\text{fm}$. For future experimental measurements of $G_A^Z(Q^2)$ and $G_T^Z(Q^2)$, we also derive the full tree-level unpolarized differential cross sections for neutrino-proton and antineutrino-proton elastic scattering in the lab frame, in hoping to provide a complementary and new perspective to unveil the nucleon spin structure by using (anti)neutrino-based facilities.}

\FullConference{
The 26th international symposium on spin physics (SPIN2025)\\
21-26 September, 2025\\
Qingdao (Tsingtao), Shandong Province, China\\}
\begin{document}
\maketitle
\section{Introduction}

Spin is a fundamental property of a particle and is of quantum nature. It is closely related to both quantum statistics for particle distributions and the Pauli exclusion principle. As spin-$\frac{1}{2}$ composite systems, nucleons (i.e. protons and neutrons) are evidently the key hadrons to study for understanding both quantum chromodynamics (QCD) and the asymmetry between matter and antimatter of the Universe, since they are the building blocks of atomic nuclei and are responsible for more than $99\%$ of the visible-matter mass of the Universe.

Internal structure information of a hadron is encoded in Lorentz-invariant functions, known as form factors (FFs). For example, the weak neutral-current (NC) axial-vector FFs of a hadron describe how the hadron reacts with $Z^0$ bosons via weak NC interactions in an elastic (anti)neutrino (or charged lepton) scattering reaction, encoding therefore clean internal axial charge and spin information of the hadron in the weak sector. Moreover, the axial FF is also closely related to the quark spin (or helicity) contribution to the hadron spin~\cite{Jaffe:1989jz,Ji:1996ek,Leader:2013jra}, playing therefore a crucial role for understanding the hadronic spin structure. When it comes to (anti)neutrino-proton or (anti)neutrino-nucleus elastic scattering~\cite{Horstkotte:1981ne,Ahrens:1986xe,MiniBooNE:2010xqw,MiniBooNE:2013dds,Ren:2022qut}, nucleon or nuclear axial-vector FFs become particularly important.

%%%%%%%%%%%%%%%%%%%%%%%%%%%%%%%%%%%%%%%%%%%%%%%%%%%%%%%%%%%%%%%%%%%%%%%%%%%%%%%%%%
\section{Axial-vector form factors and (anti)neutrino-proton elastic scattering}

For a generic spin-$\frac{1}{2}$ hadron $h$ (e.g. the proton), the hadronic matrix elements of the weak NC operator $\hat j^\mu_Z(x) \equiv \sum_q \hat{\bar q}(x) \gamma^\mu (g_V^q-g_A^q \gamma^5) \hat q(x)$ in general can be parametrized as~\cite{LlewellynSmith:1971uhs,Ohlsson:1998bk,Bernard:2001rs,Singh:2006xp,Chen:2024ksq,Chen:2026xid,Chen:2026uhf},
\begin{equation}
	\begin{aligned}\label{elements-NCO-proton}
		\langle h(p',s')|\hat j^\mu_Z(0)|h(p,s)\rangle
		= \bar u(p',s') \Gamma_Z^\mu(P,\Delta) u(p,s),
	\end{aligned}
\end{equation}
with
\begin{equation}
	\begin{aligned}\label{Vertex-Fun-Weak}
		\Gamma_Z^\mu(P,\Delta)
		&= \Gamma_V^\mu(P,\Delta) - \Gamma_A^\mu(P,\Delta),\\
		\Gamma_V^\mu(P,\Delta)
		&\equiv \gamma^\mu G_M^Z + \frac{P^\mu (G_E^Z - G_M^Z)}{M(1+\tau)} + \frac{i\Delta^\mu }{2M} G_S^Z,\\
		\Gamma^\mu_A(P,\Delta)
		&\equiv \left[ \gamma^\mu G_A^Z + \frac{\Delta^\mu}{2M} G_P^Z - \frac{\sigma^{\mu\nu} \Delta_\nu }{2M} G_T^Z \right]\gamma^5,
	\end{aligned}
\end{equation}
where $q\in\{u,d,s,c,b,t\}$, $\hat q(x)$ is the quark field operator, $g_A^{u,c,t}=\tfrac{1}{2}$, $g_A^{d,s,b}=-\tfrac{1}{2}$, $g_V^{u,c,t}=\tfrac{1}{2}-\tfrac{4}{3}\sin^2\theta_W$, $g_V^{d,s,b}=-\tfrac{1}{2}+\tfrac{2}{3}\sin^2\theta_W$, $\theta_W$ is the Weinberg angle, $P\equiv (p'+p)/2$ is the average four-momentum, $\Delta\equiv p'-p$ is the four-momentum transfer, $M$ is the mass of the hadron $h$, $Q^2\equiv -\Delta^2$, $\tau \equiv Q^2/(4M^2)$, and $u(p,s)$ denotes the Dirac spinor. Besides, $G_X^Z \equiv G_X^Z(Q^2)$ for $X=E,M,S,A,P,T$ are the weak NC FFs of the hadron, called electric, magnetic, induced scalar, axial, induced pseudoscalar and induced pseudotensor FFs, respectively. By the virtue of Hermiticity condition $\Gamma_Z^{\mu,\dag}(P,\Delta)=\gamma^0 \Gamma_Z^\mu(P,-\Delta)\gamma^0$, one can explicitly show that all these six FFs $G_X^Z(Q^2)$ are real~\cite{Chen:2024ksq,Chen:2026xid}. Furthermore, we note that any nonzero value of $G_{A,P,T}^Z(Q^2)$ implies the explicit violation of the parity symmetry~\cite{Chen:2026uhf}. Due to this reason, physically meaningful 3D root-mean-square axial radii (in the Breit frame) of any spin-$\frac{1}{2}$ targets do not exist~\cite{Chen:2024oxx,Chen:2024ksq}.

By construction, $G_{A,P,T}^Z$ can be expressed as $G_{A,P,T}^Z(Q^2) = \sum_q g_A^q G_{A,P,T}^q(Q^2)$~\cite{Chen:2024ksq,Chen:2026xid}, where $G_{A,P,T}^q(Q^2)$ are the quark-flavor axial-vector FFs; they are also accessible via measurements of polarized quark generalized parton distributions (GPDs) $\tilde H^q(x,\xi,t)$, $\tilde E^q(x,\xi,t)$ and $\tilde T^q(x,\xi,t)$ in exclusive processes using facilities such as HERA, CERN SPS, CEBAF, EIC, EicC, NICA, LHC, etc. through the first-moment sum rule~\cite{Ji:1996ek,Chen:2024ksq,Chen:2026xid}:
\begin{equation}
	\begin{aligned}\label{relation_to_GPDs}
		\begin{bmatrix}
			G_A^Z(t)\\
			G_P^Z(t)\\
			G_T^Z(t)
		\end{bmatrix}
		= \sum_q g_A^q \int_{-1}^{1} \ud x
		\begin{bmatrix}
			\tilde H^q(x,\xi,t)\\
			\tilde E^q(x,\xi,t)\\
			\tilde T^q(x,\xi,t)\\
		\end{bmatrix},
	\end{aligned}
\end{equation}
with $t \equiv \Delta^2$. The polarized quark GPDs $\tilde H^q(x,\xi,t)$, $\tilde E^q(x,\xi,t)$ and $\tilde T^q(x,\xi,t)$ are defined through the following light-front (LF) correlation function~\cite{Ji:1996ek,Chen:2026xid}
\begin{equation}
	\begin{aligned}\label{axial-vector-GPDs}
		&\int \frac{\ud z^- }{4\pi} e^{i x P^+ z^- } {_\text{LF}}\big{\langle} h(p',s') \big{|} \hat{\bar q}\left( -\tfrac{z}{2} \right) \gamma^\mu \gamma^5 \hat{\mathcal W}\left[-\tfrac{z}{2}, \tfrac{z}{2} \right] \hat{q}\left( \tfrac{z}{2} \right) \big{|} h(p,s) \big{\rangle}_\text{LF}\, \Big{|}_{z^+ = |\uvec z_\perp| = 0} \\
		&= \frac{1 }{2P^+} \bar u_\text{LF}(p',s') \! \left[ \gamma^\mu  \tilde H^q(x,\xi,t) + \frac{ \Delta^\mu }{2M} \tilde E^q(x,\xi,t) + \frac{ -\sigma^{\mu\nu} \Delta_\nu }{2M} \tilde T^q(x,\xi,t) + \cdots \right]\!\gamma^5 u_\text{LF}(p,s).
	\end{aligned}
\end{equation}
For simplicity, we have neglected the high-twist contributions that are denoted by the ellipsis.

For future measurements of $G_A^Z(Q^2)$ and $G_T^Z(Q^2)$, let us now focus on the weak NC (anti)neutrino-proton elastic scattering~\cite{Horstkotte:1981ne,Ahrens:1986xe,MiniBooNE:2010xqw,MiniBooNE:2013dds,Ren:2022qut}. Without loss of generality, we consider the following reaction:
\begin{equation}\label{vN-reaction}
	\nu_\ell(k,r) + N(p,s) \to \nu_\ell(k',r') + N(p',s'),
\end{equation}
where $\ell \in \{ e, \mu, \tau \}$, $\nu_\ell$ represents a neutrino of mass $m$, and $N \in \{p, n\}$ represents a nucleon of mass $M$, with $k^2=k'^2=m^2$ and $p^2 = p'^2 = M^2$. After some algebra and neglecting for simplicity the neutrino mass, we find that the full tree-level unpolarized differential cross sections of (anti)neutrino-proton elastic scattering (\ref{vN-reaction}) in the lab frame are explicitly obtained as~\cite{LlewellynSmith:1971uhs,Chen:2024ksq,Chen:2026xid}
\begin{equation}\label{DiffSgm-vN}
	\frac{\ud \sigma^{\pm} }{\ud Q^2}(E_\nu,Q^2) \bigg|_{\text{lab},m=0}
	= \frac{G_F^2 M^2 }{ 8\pi E_{\nu}^2 } \left(\frac{M_Z^2}{M_Z^2+Q^2 } \right)^2 \left[ A(Q^2) \pm \frac{s-u}{M^2}B(Q^2) + \frac{(s-u)^2 }{M^4} C(Q^2) \right],
\end{equation}
with
\begin{equation}
	\begin{aligned}
		A
		&\equiv 4\tau \left[ (1+\tau) (G_A^Z)^2 + \tau (G_M^Z)^2 - (G_E^Z)^2 - \tau(1+\tau) (G_T^Z)^2 \right],\\
		B
		&\equiv 4\tau G_A^Z G_M^Z,\\
		C
		&\equiv \frac{1}{4(1+\tau)} \left[ (1+\tau) (G_A^Z)^2 + \tau (G_M^Z)^2 + (G_E^Z)^2 + \tau(1+\tau) (G_T^Z)^2 \right],
	\end{aligned}
\end{equation}
where the $+$ ($-$) sign corresponds to the neutrino (antineutrino) scattering (\ref{vN-reaction}), $s\equiv (p+k)^2$ and $u\equiv (k'-p)^2$ are Mandelstam variables, $s-u=4ME_\nu-Q^2$, and we have for clarity omitted the explicit $Q^2$ dependence of these FFs $G_{E,M,A,T}^Z(Q^2)$. Last but not least, we emphasize that any experimental data of $G_A^Z(Q^2)$ and $G_T^Z(Q^2)$ obtained via (\ref{DiffSgm-vN}) in (anti)neutrino-proton elastic scattering will provide additional and useful constraints at each $Q^2$ value on the sum of quark GPDs $\tilde H^q$ and $\tilde T^q$ obtained in charged lepton-proton scattering, helping to connect the two fields~\cite{Chen:2026uhf}.

%%%%%%%%%%%%%%%%%%%%%%%%%%%%%%%%%%%%%%%%%%%%%%%%%%%%%%%%%%%%%%%%%%%%%%%%%%%%%%%%%%
\section{Relativistic 3D axial-vector four-current and spin distributions inside the proton}

Although FFs are objects defined in momentum space and extracted from experimental data involving particles with well-defined momenta, their physical interpretation actually resides in position space~\cite{Chen:2023dxp}. In this section, following our recent works~\cite{Chen:2024oxx,Chen:2024ksq}, we will present the physical density interpretation for the proton's weak NC axial-vector FFs $G_{A,P,T}^Z(Q^2)$ so as to define relativistic 3D axial-vector four-current and spin distributions inside a proton by utilizing the quantum phase-space formalism~\cite{Belitsky:2003nz,Lorce:2018egm,Lorce:2020onh,Chen:2022smg,Chen:2023dxp,Won:2022cyy,Chen:2024ksq,Lorce:2025pxt}. We emphasize that (anti)neutrino-proton elastic scattering indeed can provide useful and complementary information on the proton spin structure.

\begin{figure}[h!]
	\centering
	{\includegraphics[angle=0,scale=0.400]{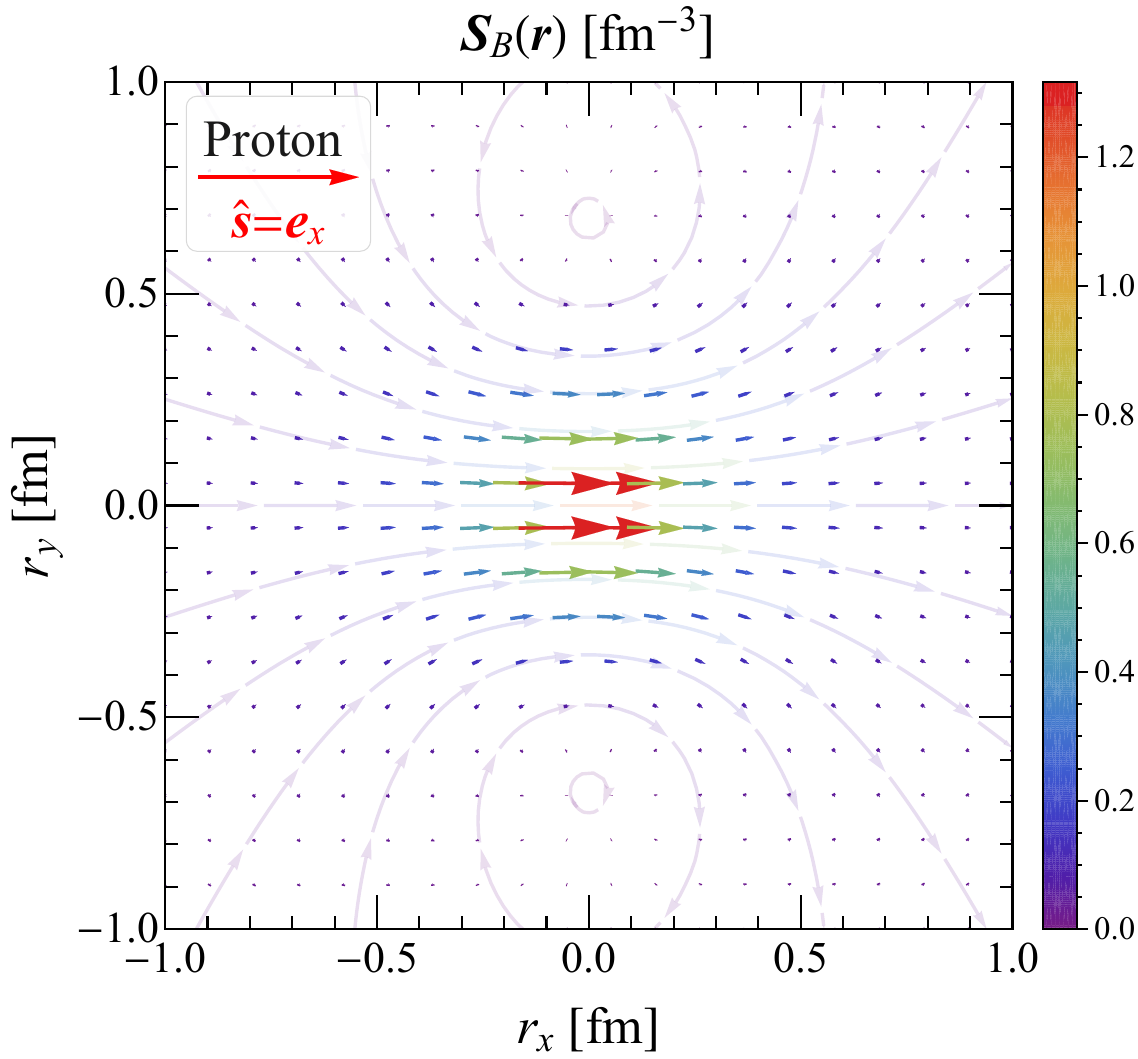}}
	\caption{Illustration of the relativistic 3D weak-neutral spin distribution $\uvec S_{B}(\uvec r)$ in the Breit frame for a transversely polarized proton in the transverse plane ($r_z=0$). See texts for more details.}
	\label{Fig_3DBFNC_Sv}
\end{figure}

In the quantum phase-space formalism, the relativistic 3D axial-vector four-current distributions inside a spin-$\frac{1}{2}$ hadron are defined in the Breit frame as~\cite{Lorce:2018egm,Lorce:2020onh,Chen:2023dxp,Chen:2024oxx,Chen:2024ksq}
\begin{equation}
	\begin{aligned}\label{BF-def}
		J_{5,B}^\mu (\uvec r)
		&\equiv \int\frac{\ud^3\Delta}{(2\pi)^3}\, e^{-i\uvec\Delta \cdot \uvec r}\, \frac{\langle p_B', s_B'| \hat j_5^\mu(0)|p_B, s_B \rangle }{2P_B^0}\bigg|_{\uvec P = \uvec 0}\\
		&= \int\frac{\ud^3\Delta}{(2\pi)^3}\, e^{-i\uvec\Delta \cdot \uvec r}\, \frac{ \bar u(p_B',s_B') }{ 2M\sqrt{1+\tau} } \left[ \gamma^\mu G_A^Z(\uvec\Delta^2) + \frac{\Delta^\mu}{2M} G_P^Z(\uvec\Delta^2) - \frac{\sigma^{\mu\nu} \Delta_\nu }{2M} G_T^Z(\uvec\Delta^2) \right] \gamma^5 u(p_B,s_B),
	\end{aligned}
\end{equation}
where $\hat j_5^\mu(x) \equiv \sum_q g_A^q \hat{\bar q}(x) \gamma^\mu \gamma^5\hat q(x)$, $\uvec r \equiv \uvec x -\uvec R$ is the distance relative to the center $\uvec R=\uvec 0$ of the system, $\uvec p_B'=-\uvec p_B = \uvec\Delta/2$, $Q^2=\uvec\Delta^2$, and $P^0_B=M\sqrt{1+\tau}$. More specifically, we have~\cite{Chen:2024oxx,Chen:2024ksq}
\begin{equation}
	\begin{aligned}\label{3DBF-distributions}
		J_{5,B}^0(\uvec r)
		&= \int\frac{\ud^3\Delta}{(2\pi)^3}\, e^{-i\uvec\Delta \cdot \uvec r}\, \frac{ i\uvec\Delta \cdot \uvec \sigma}{2M} G_T^Z(\uvec\Delta^2) = -\uvec\nabla_{\uvec r} \cdot \frac{ \uvec\sigma }{2M}\int\frac{\ud^3\Delta}{(2\pi)^3}\, e^{-i\uvec\Delta \cdot \uvec r}\, G_T^Z(\uvec\Delta^2),\\
		\uvec J_{5,B}(\uvec r)
		&= \int\frac{\ud^3\Delta}{(2\pi)^3}\, e^{-i\uvec\Delta \cdot \uvec r}\, \left\lbrace  \left[ \uvec\sigma - \frac{ \uvec\Delta (\uvec\Delta \cdot \uvec\sigma ) }{4P_B^0(P_B^0+M) } \right] G_A^Z(\uvec\Delta^2) - \frac{\uvec\Delta (\uvec\Delta \cdot \uvec\sigma ) }{ 4M P_B^0 } G_P^Z(\uvec\Delta^2) \right\rbrace,
	\end{aligned}
\end{equation}
where the explicit canonical polarization indices $s_B,\, s_B'$ are omitted for better legibility hereafter. Above results (\ref{3DBF-distributions}) clearly demonstrate that:\\
(1). The relativistic 3D axial charge distribution $J_{5,B}^0(\uvec r)$ in the Breit frame is completely described by $G_T^Z(Q^2)$ rather than by $G_A^Z(Q^2)$. Contrary to what is stated in the literature~\cite{Panteleeva:2024vdw}, we did not find any proper justification for
interpreting the quantity $R_A = \sqrt{ \langle R_A^2 \rangle } $, defined by~\cite{Meissner:1986js,Bernard:1992ys,Hill:2017wgb,MINERvA:2023avz}
\begin{equation}
	\begin{aligned}\label{naive-3DMS-axial-radius}
		\langle R_A^2 \rangle \equiv -\frac{6}{G_A^Z(0) }\frac{\ud G_A^Z(Q^2) }{\ud Q^2} \bigg|_{Q^2=0} = \frac{1}{G_A^Z(0)}\left[ -\uvec\nabla_{\uvec\Delta}^2 G_A^Z(\uvec\Delta^2) \right] \Big|_{\uvec\Delta = \uvec 0},
	\end{aligned}
\end{equation}
as the physically meaningful 3D root-mean-square axial radius of a spin-$\frac{1}{2}$ hadron~\cite{Chen:2024oxx,Chen:2024ksq}. In fact, owing the parity-odd nature of $J_{5,B}^0(\uvec r)$, the genuine 3D mean-square axial charge radius $\langle r_A^2 \rangle \equiv \frac{\int \ud^3r\, r^2\, J_{5,B}^0(\uvec r)}{\int \ud^3r\, J_{5,B}^0(\uvec r) }$ does not exist for any spin-$\frac{1}{2}$ hadron~\cite{Chen:2024ksq}.\\
(2). The axial-vector current distribution $\uvec J_{5,B}(\uvec r)$ is twice of the weak-neutral spin distribution $\uvec S_{B}(\uvec r)$. Using the QCD equation of motion, the rank-$3$ weak-neutral canonical spin tensor operator can be expressed as $\hat S^{\mu \alpha \beta } = \frac{1}{2} \epsilon^{\mu\alpha \beta \lambda} \hat j_{5\lambda}$ with $\epsilon_{0123}=+1$~\cite{Leader:2013jra}, such that $S^i \equiv \frac{1}{2} \epsilon^{ijk} S^{0jk} =  \frac{1}{2} J_5^i$, or equivalently $\uvec S = \frac{1}{2} \uvec J_5$, valid in any Lorentz frames. Hence, the relativistic 3D weak-neutral spin distribution in the Breit frame for a spin-$\frac{1}{2}$ hadron is simply given by
\begin{equation}
	\begin{aligned}
		\uvec S_{B}(\uvec r) = \frac{1}{2} \uvec J_{5,B} (\uvec r).
	\end{aligned}
\end{equation}
Based on $\uvec S_{B}(\uvec r)$, one can easily derive the following physically meaningful mean-square spin radius $\langle r_\text{spin}^2 \rangle$, defined as~\cite{Chen:2024oxx}
\begin{equation}
	\begin{aligned}\label{MS-spin-radius}
		\langle r_\text{spin}^2 \rangle \equiv \frac{ \int \ud^3r\, r^2\, \hat{\uvec s} \cdot \uvec S_{B}(\uvec r) }{\int \ud^3r\, \hat{\uvec s} \cdot \uvec S_{B}(\uvec r) } = R_A^2 + \frac{ 1 }{ 4M^2 }\left[ 1 + \frac{ 2 G_P^Z(0) }{ G_A^Z(0) } \right],
	\end{aligned}
\end{equation}
where $\hat{\uvec s} \equiv \langle s_B'| \uvec\sigma | s_B \rangle $ is an arbitrary unit spin polarization vector. Above result (\ref{MS-spin-radius}) provides an additional key motivation for the further determination of $G_P^Z(Q^2)$~\cite{Chen:2024oxx,Chen:2024ksq}. In Figure~\ref{Fig_3DBFNC_Sv}, we illustrate the relativistic 3D weak-neutral spin distribution $\uvec S_{B}(\uvec r)$ in the Breit frame for a transversely polarized (i.e. $\hat{\uvec s} = \uvec e_x$) proton in the transverse plane ($r_z=0$), by using the proton's axial-vector FFs $G_A^Z(Q^2)$ and $G_P^Z(Q^2)$ given in~\cite{Chen:2024ksq}. Numerically, we find that $r_\text{spin} = \sqrt{ \langle r_\text{spin}^2 \rangle } \approx 2.1054~\text{fm}$ and $R_A \approx 0.6510~\text{fm}$.

%%%%%%%%%%%%%%%%%%%%%%%%%%%%%%%%%%%%%%%%%%%%%%%%%%%%%%%%%%%%%%%%%%%%%%%%%%%%%%%%%%
\section{Conclusions and outlook}

Relativistic 3D weak-neutral axial-vector four-current and spin distributions inside a nucleon in terms of the nucleon's three weak neutral-current axial-vector form factors are investigated for the first time~\cite{Chen:2024oxx,Chen:2024ksq}. For future experimental measurements of $G_{A}^Z(Q^2)$ and $G_{T}^Z(Q^2)$, we also derive the full tree-level unpolarized differential cross sections for neutrino-proton and antineutrino-proton elastic scattering in the lab frame, in hoping to provide a complementary and new perspective to unveil the nucleon spin structure by using (anti)neutrino-based facilities~\cite{Chen:2026xid}.

%%%%%%%%%%%%%%%%%%%%%%%%%%%%%%%%%%%%%%%%%%%%%%%%%%%%%%%%%%%%%%%%%%%%%%%%%%%%%%%%%%
\acknowledgments

We gratefully thank Stephen~F.~Pate, Raza~S.~Sufian, Oleksandr~Tomalak, and Bo-Wen Xiao for helpful communications, and Qing Chen, Dao-Neng Gao, Feng-Kun Guo, Chueng-Ryong~Ji, Yang Li, C\'edric Lorc\'e, Qun Wang, Guang-Peng Zhang, Ren-You Zhang and Bing-Song Zou for valuable discussions. This work is supported in part by the National Natural Science Foundation of China under Grants No.~12135011 and No.~12547178, by the Strategic Priority Research Program of the Chinese Academy of Sciences under Grant No.~XDB34030102, by the Chinese Academy of
Sciences under Grant No.~YSBR-101, and by the Tsinghua University.

%%%%%%%%%%%%%%%%%%%%%%%%%%%%%%%%%%%%%%%%%%%%%%%%%%%%%%%%%%%%%%%%%%%%%%%%%%%%%%%%%%

%%%%%%%%%%%%%%%%%%%%%%%%%%%%%%%%%%%%%%%%%%%%%%%%%%%%%%%%%%%%%%%%%%%%%%%%%%%%%%%%%%

\begin{thebibliography}{99}
%%%%%%%%%%%%%%%%%%%%%%%%%%%%%%%%%%%%%%%%%%%%%%%%%%%%%%
\bibitem{Jaffe:1989jz}
R.~L.~Jaffe and A.~Manohar,
\href{https://doi.org/10.1016/0550-3213(90)90506-9}
{\emph{Nucl. Phys. B} \textbf{337}, 509-546 (1990)}.
%%%%%%%%%%%%%%%%%%%%%%%%%%%%%%%%%%%%%%%%%%%%%%%%%%%%%%
\bibitem{Ji:1996ek}
X.~D.~Ji,
\href{https://doi.org/10.1103/PhysRevLett.78.610}
{\emph{Phys. Rev. Lett.} \textbf{78}, 610 (1997)}; M.~Diehl, \href{https://doi.org/10.1016/j.physrep.2003.08.002}
{\emph{Phys. Rept.} \textbf{388}, 41 (2003)}.
%%%%%%%%%%%%%%%%%%%%%%%%%%%%%%%%%%%%%%%%%%%%%%%%%%%%%%
\bibitem{Leader:2013jra}
E.~Leader and C.~Lorc{\'e},
\href{https://doi.org/10.1016/j.physrep.2014.02.010}
{\emph{Phys. Rept.} \textbf{541}, 163-248 (2014)} [\href{https://doi.org/10.48550/arXiv.1309.4235}
{{\tt 1309.4235}}].
%%%%%%%%%%%%%%%%%%%%%%%%%%%%%%%%%%%%%%%%%%%%%%%%%%%%%%
\bibitem{Horstkotte:1981ne}
J.~Horstkotte \textit{et al.},
\href{https://doi.org/10.1103/PhysRevD.25.2743}
{\emph{Phys. Rev. D} \textbf{25}, 2743 (1982)}.
%%%%%%%%%%%%%%%%%%%%%%%%%%%%%%%%%%%%%%%%%%%%%%%%%%%%%%
\bibitem{Ahrens:1986xe}
L.~A.~Ahrens \textit{et al.},
\href{https://doi.org/10.1103/PhysRevD.35.785}
{\emph{Phys. Rev. D} \textbf{35}, 785 (1987)}.
%%%%%%%%%%%%%%%%%%%%%%%%%%%%%%%%%%%%%%%%%%%%%%%%%%%%%%
\bibitem{MiniBooNE:2010xqw}
A.~A.~Aguilar-Arevalo \textit{et al.} [MiniBooNE],
\href{https://doi.org/10.1103/PhysRevD.82.092005}
{\emph{Phys. Rev. D} \textbf{82}, 092005 (2010)} [\href{https://doi.org/10.48550/arXiv.1007.4730}
{{\tt 1007.4730}}].
%%%%%%%%%%%%%%%%%%%%%%%%%%%%%%%%%%%%%%%%%%%%%%%%%%%%%%
\bibitem{MiniBooNE:2013dds}
A.~A.~Aguilar-Arevalo \textit{et al.} [MiniBooNE],
\href{https://doi.org/10.1103/PhysRevD.91.012004}
{\emph{Phys. Rev. D} \textbf{91}, 012004 (2015)} [\href{https://doi.org/10.48550/arXiv.1309.7257}
{{\tt 1309.7257}}].
%%%%%%%%%%%%%%%%%%%%%%%%%%%%%%%%%%%%%%%%%%%%%%%%%%%%%%
\bibitem{Ren:2022qut}
L.~Ren [MicroBooNE],
\href{https://doi.org/10.22323/1.402.0205}
{\emph{PoS} \textbf{NuFact2021}, 205 (2022)}.
%%%%%%%%%%%%%%%%%%%%%%%%%%%%%%%%%%%%%%%%%%%%%%%%%%%%%%
\bibitem{LlewellynSmith:1971uhs}
C.~H.~Llewellyn Smith,
\href{https://doi.org/10.1016/0370-1573(72)90010-5}
{\emph{Phys. Rept.} \textbf{3}, 261-379 (1972)}.
%%%%%%%%%%%%%%%%%%%%%%%%%%%%%%%%%%%%%%%%%%%%%%%%%%%%%%
\bibitem{Ohlsson:1998bk}
T.~Ohlsson and H.~Snellman,
\href{https://doi.org/10.1007/s100529800908}
{\emph{Eur. Phys. J. C} \textbf{6}, 285-296 (1999)} [\href{https://doi.org/10.48550/arXiv.hep-ph/9803490}
{{\tt hep-ph/9803490}}].
%%%%%%%%%%%%%%%%%%%%%%%%%%%%%%%%%%%%%%%%%%%%%%%%%%%%%%
\bibitem{Bernard:2001rs}
V.~Bernard, L.~Elouadrhiri and U.~Meissner,
\href{https://doi.org/10.1088/0954-3899/28/1/201}
{\emph{J. Phys. G} \textbf{28}, R1-R35 (2002)} [\href{https://doi.org/10.48550/arXiv.hep-ph/0107088}
{{\tt hep-ph/0107088}}].
%%%%%%%%%%%%%%%%%%%%%%%%%%%%%%%%%%%%%%%%%%%%%%%%%%%%%%
\bibitem{Singh:2006xp}
S.~K.~Singh and M.~J.~Vicente Vacas,
\href{https://doi.org/10.1103/PhysRevD.74.053009}
{\emph{Phys. Rev. D} \textbf{74}, 053009 (2006)} [\href{https://doi.org/10.48550/arXiv.hep-ph/0606235}
{{\tt hep-ph/0606235}}].
%%%%%%%%%%%%%%%%%%%%%%%%%%%%%%%%%%%%%%%%%%%%%%%%%%%%%%
\bibitem{Chen:2024ksq}
Y.~Chen,
\href{https://doi.org/10.1007/JHEP04(2025)132}
{\emph{JHEP} \textbf{04}, 132 (2025)} [\href{https://doi.org/10.48550/arXiv.2411.12521}
{{\tt 2411.12521}}]; ibid.,
[\href{https://arxiv.org/abs/2411.12521v2}
{{\tt 2411.12521v2}}].	
%%%%%%%%%%%%%%%%%%%%%%%%%%%%%%%%%%%%%%%%%%%%%%%%%%%%%%
\bibitem{Chen:2026xid}
Y.~Chen, O.~Tomalak and B.-S.~Zou, [\href{https://doi.org/10.48550/arXiv.2607.15050}
{{\tt 2607.15050}}]; ibid., To appear.
%%%%%%%%%%%%%%%%%%%%%%%%%%%%%%%%%%%%%%%%%%%%%%%%%%%%%%
\bibitem{Chen:2026uhf}
Y.~Chen, Q.~Chen, F.-K.~Guo, Q.~Wang and B.-S.~Zou, To appear.
%%%%%%%%%%%%%%%%%%%%%%%%%%%%%%%%%%%%%%%%%%%%%%%%%%%%%%
\bibitem{Chen:2023dxp}
Y.~Chen and C.~Lorc\'e,
\href{https://doi.org/10.1103/PhysRevD.107.096003}
{\emph{Phys. Rev. D} \textbf{107}, 096003 (2023)} [\href{https://doi.org/10.48550/arXiv.2302.04672}
{{\tt 2302.04672}}].
%%%%%%%%%%%%%%%%%%%%%%%%%%%%%%%%%%%%%%%%%%%%%%%%%%%%%%
\bibitem{Chen:2024oxx}
Y.~Chen, Y.~Li, C.~Lorc{\'e} and Q.~Wang,
\href{https://doi.org/10.1103/PhysRevD.110.L091503}
{\emph{Phys. Rev. D} \textbf{110}, L091503 (2024)} [\href{https://doi.org/10.48550/arXiv.2405.12943}
{{\tt 2405.12943}}].
%%%%%%%%%%%%%%%%%%%%%%%%%%%%%%%%%%%%%%%%%%%%%%%%%%%%%%
\bibitem{Belitsky:2003nz}
A.~V.~Belitsky, X.~d.~Ji and F.~Yuan,
\href{https://doi.org/10.1103/PhysRevD.69.074014}
{\emph{Phys. Rev. D} \textbf{69}, 074014 (2004)} [\href{https://doi.org/10.48550/arXiv.hep-ph/0307383}
{{\tt hep-ph/0307383}}].
%%%%%%%%%%%%%%%%%%%%%%%%%%%%%%%%%%%%%%%%%%%%%%%%%%%%%%
\bibitem{Lorce:2018egm}
C.~Lorc{\'e}, H.~Moutarde and A.~P.~Trawi{\'n}ski,
\href{https://doi.org/10.1140/epjc/s10052-019-6572-3}
{\emph{Eur. Phys. J. C} \textbf{79}, 89 (2019)} [\href{https://doi.org/10.48550/arXiv.1810.09837}
{{\tt 1810.09837}}].
%%%%%%%%%%%%%%%%%%%%%%%%%%%%%%%%%%%%%%%%%%%%%%%%%%%%%%
\bibitem{Lorce:2020onh}
C.~Lorc{\'e},
\href{https://doi.org/10.1103/PhysRevLett.125.232002}
{\emph{Phys. Rev. Lett.} \textbf{125}, 232002 (2020)} [\href{https://doi.org/10.48550/arXiv.2007.05318}
{{\tt 2007.05318}}].
%%%%%%%%%%%%%%%%%%%%%%%%%%%%%%%%%%%%%%%%%%%%%%%%%%%%%%	
\bibitem{Chen:2022smg}
Y.~Chen and C.~Lorc\'e,
\href{https://doi.org/10.1103/PhysRevD.106.116024}
{\emph{Phys. Rev. D} \textbf{106}, 116024 (2022)} [\href{https://doi.org/10.48550/arXiv.2210.02908}
{{\tt 2210.02908}}].
%%%%%%%%%%%%%%%%%%%%%%%%%%%%%%%%%%%%%%%%%%%%%%%%%%%%%%
\bibitem{Won:2022cyy}
H.~Y.~Won, H.~C.~Kim and J.~Y.~Kim,
\href{https://doi.org/10.1103/PhysRevD.106.114009}
{\emph{Phys. Rev. D} \textbf{106}, 114009 (2022)} [\href{https://doi.org/10.48550/arXiv.2210.03320}
{{\tt 2210.03320}}].
%%%%%%%%%%%%%%%%%%%%%%%%%%%%%%%%%%%%%%%%%%%%%%%%%%%%%%
\bibitem{Lorce:2025pxt}
C.~Lorc{\'e}, A.~Mukherjee, R.~Singh and H.~Y.~Won,
\href{https://doi.org/10.1016/j.physletb.2025.139792}
{\emph{Phys. Lett. B} \textbf{868}, 139792 (2025)}.
%%%%%%%%%%%%%%%%%%%%%%%%%%%%%%%%%%%%%%%%%%%%%%%%%%%%%%
\bibitem{Panteleeva:2024vdw}
J.~Y.~Panteleeva, E.~Epelbaum, J.~Gegelia and U.~G.~Mei{\ss}ner,
[\href{https://doi.org/10.48550/arXiv.2412.05050}
{{\tt 2412.05050}}].
%%%%%%%%%%%%%%%%%%%%%%%%%%%%%%%%%%%%%%%%%%%%%%%%%%%%%%
\bibitem{Meissner:1986js}
U.~G.~Meissner, N.~Kaiser and W.~Weise,
\href{https://doi.org/10.1016/0375-9474(87)90463-5}
{\emph{Nucl. Phys. A} \textbf{466}, 685-723 (1987)}.
%%%%%%%%%%%%%%%%%%%%%%%%%%%%%%%%%%%%%%%%%%%%%%%%%%%%%%
\bibitem{Bernard:1992ys}
V.~Bernard, N.~Kaiser and U.~G.~Meissner,
\href{https://doi.org/10.1103/PhysRevLett.69.1877}
{\emph{Phys. Rev. Lett.} \textbf{69}, 1877-1879 (1992)}.
%%%%%%%%%%%%%%%%%%%%%%%%%%%%%%%%%%%%%%%%%%%%%%%%%%%%%%%
\bibitem{Hill:2017wgb}
R.~J.~Hill, P.~Kammel, W.~J.~Marciano and A.~Sirlin,
\href{https://doi.org/10.1088/1361-6633/aac190}
{\emph{Rept. Prog. Phys.} \textbf{81}, 096301 (2018)}.
%%%%%%%%%%%%%%%%%%%%%%%%%%%%%%%%%%%%%%%%%%%%%%%%%%%%%%
\bibitem{MINERvA:2023avz}
T.~Cai \textit{et al.} [MINERvA],
\href{https://doi.org/10.1038/s41586-022-05478-3}
{\emph{Nature} \textbf{614}, 48-53 (2023)}.
%%%%%%%%%%%%%%%%%%%%%%%%%%%%%%%%%%%%%%%%%%%%%%%%%%%%%%
%\bibitem{Wigner:1932eb}
%E.~P.~Wigner,
%\href{https://doi.org/10.1103/PhysRev.40.749}
%{\emph{Phys. Rev.} \textbf{40}, 749-760 (1932)}.
%%%%%%%%%%%%%%%%%%%%%%%%%%%%%%%%%%%%%%%%%%%%%%%%%%%%%%%
%\bibitem{Hillery:1983ms}
%M.~Hillery, R.~F.~O'Connell, M.~O.~Scully and E.~P.~Wigner,
%\href{https://doi.org/10.1016/0370-1573(84)90160-1}
%{\emph{Phys. Rept.} \textbf{106}, 121-167 (1984)}.
%%%%%%%%%%%%%%%%%%%%%%%%%%%%%%%%%%%%%%%%%%%%%%%%%%%%%%%
%\bibitem{Bialynicki-Birula:1991jwl}
%I.~Bialynicki-Birula, P.~Gornicki and J.~Rafelski,
%\href{https://doi.org/10.1103/PhysRevD.44.1825}
%{\emph{Phys. Rev. D} \textbf{44}, 1825-1835 (1991)}.
%%%%%%%%%%%%%%%%%%%%%%%%%%%%%%%%%%%%%%%%%%%%%%%%%%%%%%%
%\bibitem{Lorce:2018zpf}
%C.~Lorc{\'e},
%\href{https://doi.org/10.1140/epjc/s10052-018-6249-3}
%{\emph{Eur. Phys. J. C} \textbf{78}, 785 (2018)} [\href{https://doi.org/10.48550/arXiv.1805.05284}
%{{\tt 1805.05284}}].
%%%%%%%%%%%%%%%%%%%%%%%%%%%%%%%%%%%%%%%%%%%%%%%%%%%%%%%
%\bibitem{Lorce:2021gxs}
%C.~Lorc{\'e},
%\href{https://doi.org/10.1140/epjc/s10052-021-09207-4}
%{\emph{Eur. Phys. J. C} \textbf{81}, 413 (2021)} [\href{https://doi.org/10.48550/arXiv.1805.05284}
%{{\tt 2103.10100}}].
%%%%%%%%%%%%%%%%%%%%%%%%%%%%%%%%%%%%%%%%%%%%%%%%%%%%%%%
%\bibitem{Lorce:2022cle}
%C.~Lorc{\'e}, P.~Schweitzer and K.~Tezgin,
%\href{https://doi.org/10.1103/PhysRevD.106.014012}
%{\emph{Phys. Rev. D} \textbf{106}, 014012 (2022)} [\href{https://doi.org/10.48550/arXiv.2202.01192}
%{{\tt 2202.01192}}].
%%%%%%%%%%%%%%%%%%%%%%%%%%%%%%%%%%%%%%%%%%%%%%%%%%%%%%%
%\bibitem{Lorce:2022jyi}
%C.~Lorc{\'e} and P.~Wang,
%\href{https://doi.org/10.1103/PhysRevD.105.096032}
%{\emph{Phys. Rev. D} \textbf{105}, 096032 (2022)} [\href{https://doi.org/10.48550/arXiv.2204.01465}
%{{\tt 2204.01465}}].
%%%%%%%%%%%%%%%%%%%%%%%%%%%%%%%%%%%%%%%%%%%%%%%%%%%%%%%
%\bibitem{Kim:2021kum}
%J.~Y.~Kim and H.~C.~Kim,
%\href{https://doi.org/10.1103/PhysRevD.104.074003}
%{\emph{Phys. Rev. D} \textbf{104}, 074003 (2021)} [\href{https://doi.org/10.48550/arXiv.2204.01465}
%{{\tt 2106.10986}}].
%%%%%%%%%%%%%%%%%%%%%%%%%%%%%%%%%%%%%%%%%%%%%%%%%%%%%%%
%\bibitem{Kim:2022bia}
%J.~Y.~Kim,
%\href{https://doi.org/10.1103/PhysRevD.106.014022}
%{\emph{Phys. Rev. D} \textbf{106}, 014022 (2022)} [\href{https://doi.org/10.48550/arXiv.2204.08248}
%{{\tt 2204.08248}}].
%%%%%%%%%%%%%%%%%%%%%%%%%%%%%%%%%%%%%%%%%%%%%%%%%%%%%%%
%\bibitem{Hong:2023tkv}
%K.~H.~Hong, J.~Y.~Kim and H.~C.~Kim,
%\href{https://doi.org/10.1103/PhysRevD.107.074004}
%{\emph{Phys. Rev. D} \textbf{107}, 074004 (2023)} [\href{https://doi.org/10.48550/arXiv.2301.09267}
%{{\tt 2301.09267}}].
%%%%%%%%%%%%%%%%%%%%%%%%%%%%%%%%%%%%%%%%%%%%%%%%%%%%%%%
%\bibitem{Won:2025dgc}
%H.~Y.~Won and C.~Lorc{\'e},
%\href{https://doi.org/10.1103/PhysRevD.106.116024}
%{\emph{Phys. Rev. D} \textbf{111}, 094021 (2025)} [\href{https://doi.org/10.48550/arXiv.2503.07382}
%{{\tt 2503.07382}}].
%%%%%%%%%%%%%%%%%%%%%%%%%%%%%%%%%%%%%%%%%%%%%%%%%%%%%%%
%\bibitem{Lorce:2025pxt}
%C.~Lorc{\'e}, A.~Mukherjee, R.~Singh and H.~Y.~Won,
%\href{https://doi.org/10.1016/j.physletb.2025.139792}
%{\emph{Phys. Lett. B} \textbf{868}, 139792 (2025)} [\href{https://doi.org/10.48550/arXiv.2505.20468}
%{{\tt 2505.20468}}].
%%%%%%%%%%%%%%%%%%%%%%%%%%%%%%%%%%%%%%%%%%%%%%%%%%%%%%%
%
\end{thebibliography}
\end{document}